\documentclass[aps,twocolumn,showpacs,amssymb,prl]{revtex4-1}
\usepackage{amsmath}
\usepackage{hyperref}
\usepackage{graphicx}
\usepackage{bbm}
\usepackage{color}

\usepackage{bm}

\usepackage{mathtools}

\newcommand{\fl}{\textcolor{black}}

\allowdisplaybreaks

\begin{document}
\title{Finite-temperature dynamic structure factor of the spin-1 ${XXZ}$ chain with single-ion anisotropy}

\author{Florian Lange}
\author{Satoshi Ejima}
\author{Holger Fehske}
\affiliation{Institut f{\"u}r Physik,
             Ernst-Moritz-Arndt-Universit{\"a}t Greifswald,
             D-17489 Greifswald,
             Germany}

\begin{abstract}
\fl{Improving matrix-product state techniques based on the purification of the density matrix, we are able to accurately calculate the finite-temperature dynamic response of the infinite spin-1 $XXZ$ chain with single-ion anisotropy in the Haldane, large-$D$ and antiferromagnetic phases.}  Distinct thermally activated scattering processes make a significant contribution to the spectral weight in all cases. In the Haldane phase intraband magnon scattering is prominent, and the onsite anisotropy causes the magnon to split into singlet and doublet branches.  In the large-$D$ phase response, the intraband signal is separated from an exciton-antiexciton continuum.  In the antiferromagnetic phase, holons are the lowest-lying excitations, with a gap that closes at the transition to the Haldane state.  At finite temperatures, scattering between domain-wall excitations becomes especially important and strongly enhances the spectral weight for momentum transfer $\pi$.
\end{abstract}

\maketitle

The spin-$S$ $XXZ$ Heisenberg chain, defined by the Hamilton operator
\begin{equation}
\hat{H}_{XXZ} = \sum_j\left[ \frac{J}{2}(\hat{S}_j^+\hat{S}_{j+1}^- + \hat{S}_j^-\hat{S}_{j+1}^+) + J_z \hat{S}_j^z\hat{S}_{j+1}^z \right] \;,
\end{equation}
is perhaps the most fundamental model in the study of low-dimensional magnetism. 
Here, the experimentally quite often realized and regarding the nature of the excitations very different cases $S=1/2$ and $S=1$ are of particular importance. \fl{Haldane's conjecture~\cite{Ha83} states that, at the isotropic point, the ground state of a chain with integer spin is gapped while that of a half-integer spin chain is gapless. Motivated by this, there has been a continued interest in the distinct properties of the spin-1 chain. }
Unlike its spin-1/2 counterpart, however, for which numerous exact results can be obtained via the Bethe ansatz, the spin-1 chain is not integrable and one often has to rely on numerical calculations. 
Nevertheless, the ground-state phase diagram of the spin-1 chain is now well-established. For an antiferromagnetic interaction $(J_z>0)$ and when taking an additional single-ion anisotropy into account, the model exhibits Haldane, large-$D$, and antiferromagnetic (N\'eel) phases.  
These phases are realized by different compounds with Ni$^{2+}$-ions, opening up the possibility directly to compare the theoretical predictions with experimental data. 
Examples are Ni(C$_2$H$_8$N$_2$)$_2$NO$_2$(ClO$_4$) (the so-called NENP)~\cite{0295-5075-3-8-013,PhysRevB.50.9174} and SrNi$_2$V$_2$O$_8$~\cite{PhysRevB.87.224423,HaldaneChainMaterials} for the Haldane phase, NiCl$_2$−4SC(NH$_2$)$_2$ (DTN)~\cite{PhysRevLett.96.077204,PhysRevLett.98.047205} for the large-$D$ phase, and $\rm NiCl_3C_6H_5CH_2CH_2NH_3$~\cite{Lipps2017a} for the antiferromagnetic phase. 
Inelastic neutron scattering provides maybe the most comprehensive
experimental characterization of such materials. In this case the measured quantity
is the dynamic spin structure factor which contains detailed information
about the systems' excitation spectrum. 

From the theory side, a very reliable calculation of the magnetic response of one-dimensional spin systems can be performed, at zero temperature, by means of the numerical density-matrix renormalization group technique~\cite{White92,PhysRevB.77.134437}.
However, to more closely approximate the conditions in real experiments, it is desirable to take finite-temperature effects into account, such as the shift and broadening of spectral lines or the intraband scattering recently predicted for the Haldane chain with $J_z/J=1$~\cite{Goettinger}. 
A standard approach for the calculation of finite-temperature dynamics is based on evolving the purification of the density matrix in real time~\cite{Suzuki85,MPSPurification}. The main limitation of this method is the reachable time scale because of the entanglement growth out of equilibrium. A partial remedy for this is given by using time-translation invariance~\cite{TimeEvolutionTwoStates} and a backwards time-evolution on the auxiliary sites~\cite{ExtendingRange}. 

In this Rapid Communication, we combine these techniques with the infinite boundary conditions  (IBCs) originally introduced for zero-temperature calculations~\cite{InfiniteBoundary1,InfiniteBoundary2,InfiniteBoundary3}, to obtain the finite-temperature, momentum- and energy-resolved spin structure factor of the anisotropic spin-1 chain  directly in the thermodynamic limit. 
An improved scheme for the evaluation of the time-dependent correlation functions thereby allows us to significantly reduce the numerical effort when exploiting time-translation invariance. 

Hereinafter, we will first recapitulate the main previous results for the antiferromagnetic spin-1 chain with single-ion anisotropy. Then our numerical approach will be outlined, and finally we will present and discuss our findings for the dynamic spin structure factor in  three different parameter regimes, corresponding to the Haldane, large-$D$ and antiferromagnetic quantum phases. 

The Hamilton operator of the spin-1 $XXZ$ chain with single-ion anisotropy $D$ is
\begin{align}
\hat{H} &= \hat{H}_{XXZ} + D \sum_j (\hat{S}_j^z)^2 \; .
\label{model}
\end{align}
Assuming a positive exchange parameter $J>0$,  the ground-state phase diagram of the model~(\ref{model}) for $J_z/J>0$ consists of three gapped phases~\cite{PhysRevB.67.104401}. 
At the isotropic point ($D=0$, $J_z/J=1$), the ground state belongs to
the symmetry-protected topological Haldane 
phase~\cite{PhysRevB.80.155131,PhysRevB.81.064439}. 
A transition to the topologically trivial large-$D$ phase that includes the product state with $S^z=0$ at every site takes place for strong onsite positive anisotropy $D/J$. 
Lastly, a long-range ordered antiferromagnetic phase exists at negative $D/J$ or exchange anisotropy $J_z > J$.

Tackling~\eqref{model} at finite temperatures $T=1/\beta$, within the so-called purification method, the density matrix $\rho$ of the system is regarded as the reduced density matrix of a pure state $|\psi\rangle$ in an enlarged Hilbert space with twice as many sites, $\rho = \text{Tr}_Q |\psi \rangle \langle \psi \rangle$, where trace is taken over the space $Q$ spanned by the auxiliary sites. 
To obtain the equilibrium density matrix at $T$, 
 one first constructs a matrix-product state (MPS) representation of a state
$|\psi_\infty \rangle$ corresponding to the infinite-temperature density
matrix and then carries out an imaginary time evolution 
$|\psi_\beta \rangle = e^{-\beta \hat{H}/2} |\psi_\infty \rangle$ on the physical subsystem. 
A possible choice for $|\psi_\infty \rangle$ in the grand canonical ensemble is a state where each physical site is in a maximally entangled state with an auxiliary site. 
When the physical and auxiliary sites are arranged alternately, such a
state has a simple MPS representation that can be easily constructed. 
Then, for any nearest-neighbor Hamiltonian, the time evolution can be carried out with, for example, a Suzuki-Trotter decomposition and swap gates~\cite{SwapGates}. 

To avoid boundary effects, the purification method can be applied
directly in the thermodynamic limit by using infinite MPS (iMPS) that
are invariant under the translation by a unit cell. This also reduces
the number of MPS parameters since only a small unit cell is needed.
For the time evolution, one can employ the infinite 
time-evolving block decimation (TEBD) method~\cite{PhysRevLett.91.147902,iTEBD}. 
However, since the imaginary time evolution is not unitary, the canonical form 
of the iMPS is lost after each time step which leads to a rapidly growing error due to large truncations. One should therefore make use of a reorthogonalization procedure~\cite{iTEBDBeyondUnitary} to restore the canonical form. 

Dynamic properties can be calculated similarly to the $T=0$ case by switching to real-time evolution, but a fast growth of the entanglement usually restricts the simulations to short time scales.  
Several methods have been devised to extend the range of the simulations. 
A significant improvement is achieved evolving the auxiliary system in reverse time to slow down the entanglement  growth~\cite{ExtendingRange}. 
Additionally, time-translation invariance can be used to spread the time evolution to two MPS and increase the simulated time approximately by a factor of 2~\cite{TimeEvolutionTwoStates,ExtendingRange}. 
\begin{figure}[tb!]
\centering
\includegraphics[width=0.47\textwidth]{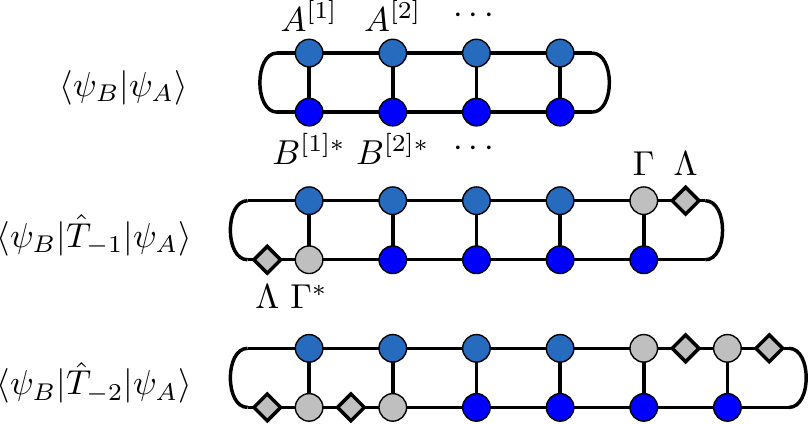}
\caption{
 Graphical representation of Eq.~\eqref{IBCcorrfunc}. The blue
 symbols represent the tensors in the finite window that distinguishes
 between $|\psi_A\rangle$ and $|\psi_B\rangle$ while the gray tensors
 represent the iMPS unit cell. 
} 
\label{figIBC}
\end{figure}
In equilibrium, the reverse time evolution on the auxiliary system completely cancels the effect of the physical time evolution for any inverse temperature $\beta$. 
When a local perturbation is applied to $|\psi_\beta\rangle$, only the tensors of the MPS in the region over which the perturbation has spread need to be updated during the time evolution. 
It is therefore possible to use IBCs to avoid finite-size effects in the calculation of dynamic correlation functions. 
IBCs are also advantageous when calculating correlation functions by using time translation invariance. In that case, both operators are fixed so that a separate simulation would be necessary for each distance if open boundary conditions are used. For IBCs, however, we can exploit the spatial translation invariance to shift both states relative to each other and obtain the correlation function at arbitrary distance.
An MPS with IBCs can be written as 
\begin{align}
|\psi \rangle &= \sum_{\bm{\sigma}} ...\Lambda \Gamma^{\sigma_0} A^{[1]\sigma_1}...A^{[N]\sigma_N} \Gamma^{\sigma_{N+1}}\Lambda ...|\bm{\sigma}\rangle \; ,
\end{align}
where $\sigma_j$ labels the basis states of the local Hilbert space at site $j$. The infinitely repeated iMPS unit cell is defined by $\Gamma$ and $\Lambda$, and only the tensors $A^{[j]}$ in a finite window are updated during the time evolution. 
For two different states $|\psi_A\rangle$ and $|\psi_B\rangle$ with tensors $A^{[j]}$ and $B^{[j]}$, respectively, and the same iMPS unit cell, we have

\begin{align}
\langle \psi_B | \hat{T}_{-r} | &\psi_A  \rangle  = \sum_{\{\sigma\}} \text{Tr}\left( A^{[1]\sigma_1}...A^{[N]\sigma_N} \Big(\prod_{d=1}^r \Gamma^{\sigma_{N+d}}\Lambda \Big) \right. \nonumber \\
 & \left. B^{[N]\sigma_{N+r} \dagger}...B^{[1]\sigma_{r+1}\dagger} \Big(\prod_{d=1}^r \Gamma^{\sigma_{r+1-d}\dagger} \Lambda \Big) \right),
\label{IBCcorrfunc}
\end{align}
where $\hat{T}_r$ is the translation operator for a shift by $r$ sites and $N$ is the number of sites in the window. Graphically, this can be represented as shown in Fig.~\ref{figIBC}.  
To calculate the dynamic correlation function for some operator $\hat{O}$, one can identify $|\psi_A\rangle = e^{-i(t/2)\hat{H}}\hat{O}_j|\psi_\beta\rangle$ and $|\psi_B\rangle = e^{+i(t/2)\hat{H}}\hat{O}_j|\psi_\beta\rangle$ so that Eq.~\eqref{IBCcorrfunc} gives $\langle \hat{O}_{j+r}^{\dagger}(t) \hat{O}_j(0) \rangle$, provided one also applies the auxiliary time evolution. In our simulations, the expectation values are taken in the grand-canonical ensemble. 
\fl{While we restrict ourselves to gapped phases, the purification method can be applied to gapless phases as well. In that case, the bond dimension required to approximate the equilibrium density matrix with a fixed accuracy would scale polynomially with the inverse temperature $\beta$ instead of saturating at large $\beta$ as for gapped phases~\cite{PurificationEntanglement}.}

The longitudinal and transversal dynamic spin structures factors we are interested in are defined as 
\begin{align}
S^{zz}(k,\omega) &= \int_{-\infty}^{\infty} dt \sum_r e^{i(kr-\omega t)} \langle \hat{S}_{j+r}^z(t) \hat{S}_j^z(0) \rangle \,, \label{Szz} \\
S^{+-}(k,\omega) &= \frac{1}{2}\int_{-\infty}^{\infty} dt \sum_r e^{i(kr-\omega t)} \langle \hat{S}_{j+r}^+(t) \hat{S}_j^-(0) \rangle \,.
\label{Spm}
\end{align}
\fl{We calculate the time-dependent correlation functions in Eqs.~\eqref{Szz} and \eqref{Spm} 
with the method described above and, to reach a higher resolution, extrapolate the data to larger times using linear prediction~\cite{PhysRevB.77.134437}.} 
\fl{The MPS simulations usually take a couple of days to finish on a modern cluster when using a parallel TEBD implementation. }

\begin{figure}[!bt]
\centering
\includegraphics[width=0.47\textwidth]{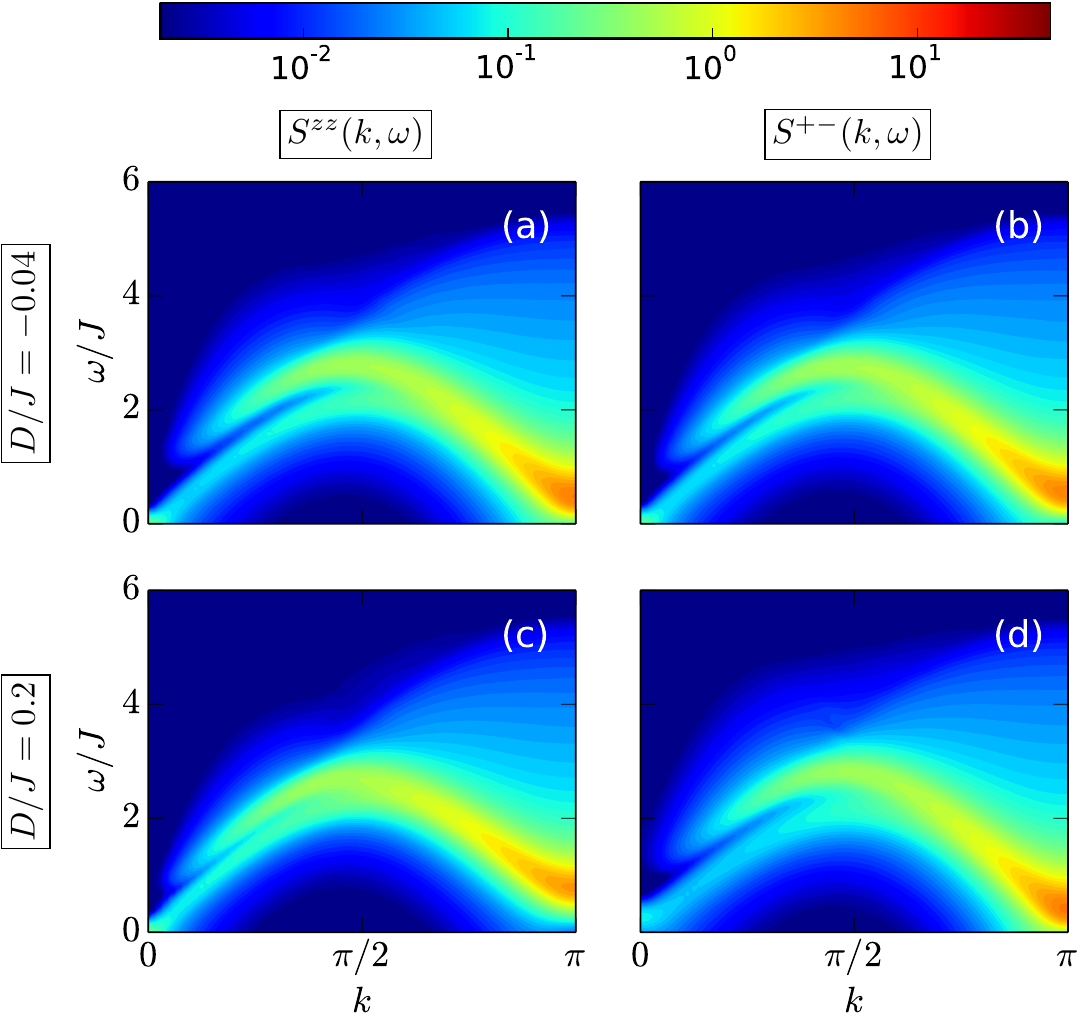}
\caption{Finite-temperature dynamic structure factor in the Haldane phase ($J_z=J$) in units of \fl{$J^{-1}$}. The temperature $T/J=0.4$; the onsite anisotropy $D/J=-0.04$  
 in (a),  (b) and $D/J=0.2$ in (c), (d). All spectral functions are convoluted with a \fl{Gaussian} of width $0.1J$. 
 }
\label{SkwHaldane}
\end{figure}

For the Haldane phase, we assume an isotropic exchange $(J_z/J=1$) and two
realistic values of the single-ion anisotropy, $D/J=-0.04$ and $D/J=0.2$, 
corresponding to the compounds 
SrNi$_2$V$_2$O$_8$~\cite{HaldaneChainMaterials}
and NENP~\cite{Delica1991}, respectively. 
Since these values are close to the isotropic point already studied in Ref.~\cite{Goettinger}, we restrict ourselves to a single intermediate temperature $T/J=0.4$. 
Figure~\ref{SkwHaldane} gives the results for the dynamic structure
factors~\eqref{Szz} and \eqref{Spm} \fl{(see also Ref.~\cite{SuppMat} for constant-momentum cuts)}. 
The onsite anisotropy causes the magnon to split into a singlet branch ($S^z=0$) and a doublet branch ($S^z=\pm1$), which show up in $S^{zz}(k,\omega)$ and $S^{+-}(k,\omega)$, respectively. For positive $D$, the singlet gap is larger than the doublet gap, while the situation is reversed for negative $D$. 
At finite temperature, there is an additional spectral weight below the magnon bands, which in the longitudinal (transversal) structure factor is caused by intraband (interband) scattering. The splitting of the magnon branch shifts the position of the interband signal in $S^{+-}(k,\omega)$ compared to the intraband response seen in $S^{zz}(k,\omega)$, so that the spectral weight for zero momentum transfer is centered at a small finite energy. 
For the considered $D$-values we find only a small effect
of the anisotropy and essentially reproduce the result of
Ref.~\cite{Goettinger} with the system size $L=32$ and open boundary
conditions. 
Note, however, that the edge-state modes of Ref.~\cite{Goettinger} are absent 
because our simulations are done in the thermodynamic limit. 

\begin{figure}[!bt]
\centering
\includegraphics[width=0.48\textwidth]{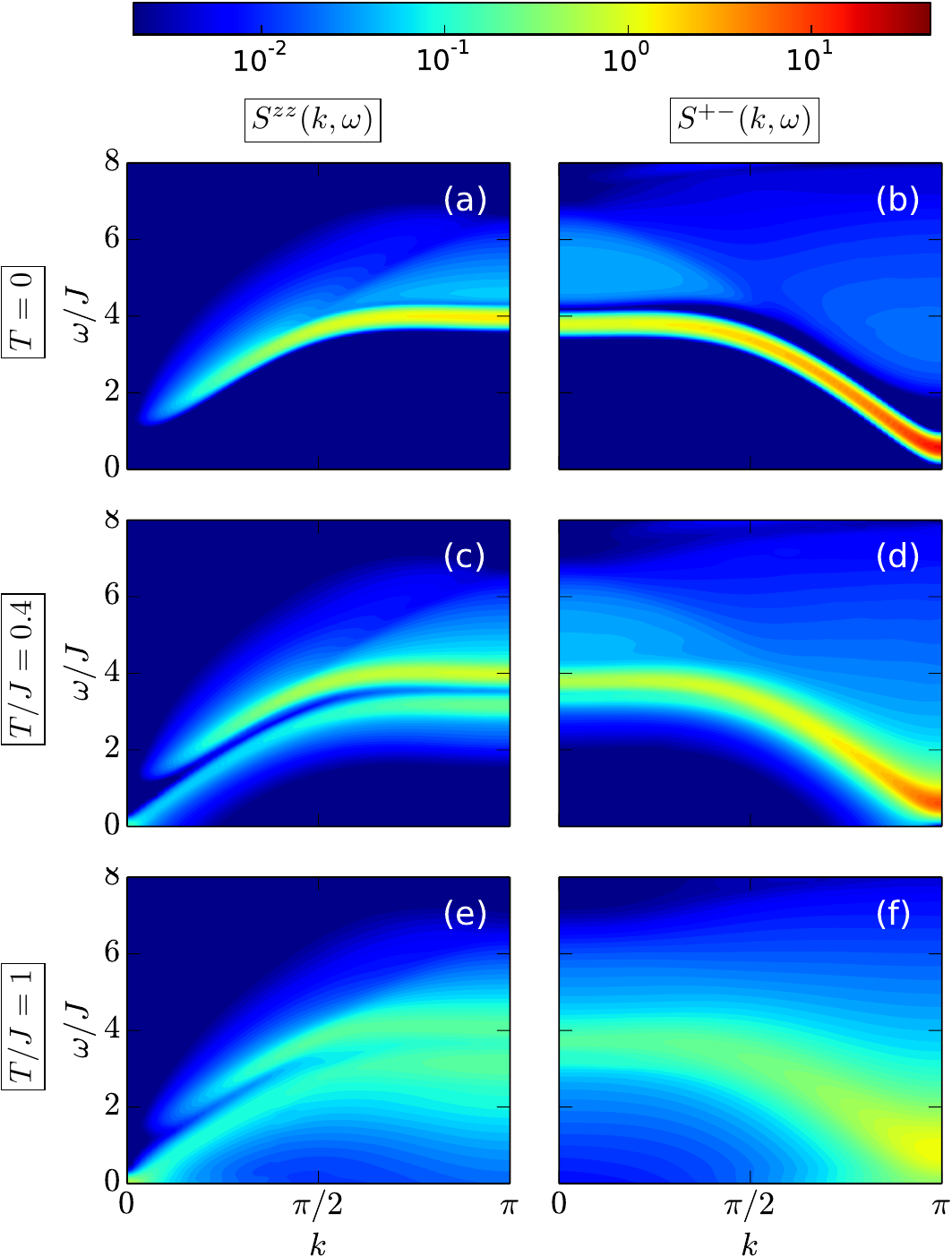}
\caption{Dynamic spin structure factor in the large-$D$ phase for $D/J=2$. 
 Again $J_z/J=1$.  Zero-temperature
 data (a), (b) obtained by pure-state MPS techniques are contrasted with the results for $T/J=0.4$ (c), (d) and $1$ (e), (f).
 }
\label{SkwLargeD}
\end{figure}

We now choose an anisotropy $D/J=2$ strong enough for the system to be in the topologically trivial large-$D$ phase (see Fig.~\ref{SkwLargeD}). 
The lowest-lying excitations in the large-$D$ phase can be viewed as single up or down spins that move in a background of sites with $S^z=0$. These excitations have been called excitons and antiexcitons~\cite{BoundState}. 
At zero temperature, the longitudinal structure factor $S^{zz}(k,\omega)$ consists of an exciton-antiexciton continuum and possibly a bound state due to the attractive interaction between opposite spins~\cite{BoundState}. For the parameters taken in Fig.~\ref{SkwLargeD}, a bound state occurs at momenta $k \gtrsim \pi/2$. 
When the temperature is increased, the dynamic structure factor broadens and the contributions of the bound state and the continuum become indistinguishable.
Similar to the thermal intraband magnon scattering in the Haldane phase, intraband scattering of excitons and antiexcitons at finite temperature produces additional spectral weight at low energies that is separated from the exciton-antiexciton continuum. 
When $D/J$ is lowered, the single-exciton gap decreases which results in a smaller distance between the intraband-scattering peak and the exciton-antiexciton continuum. 
In the zero-temperature transversal structure factor $S^{+-}(k,\omega)$, most of the spectral weight is concentrated in the single-exciton branch that lies below the three-particle continuum. 
At finite temperature, the single-exciton line broadens and eventually merges with the continuum. 
Since only matrix elements between states whose total $S^z$ differ by one contribute to $S^{+-}(k,\omega)$, no intraband scattering is observed in the transversal structure factor. For small momenta $k\approx 0$, however, an additional peak appears slightly below the single-exciton line that is likely caused by transitions between excitons and exciton-antiexciton bound states.

For the dynamic magnetic response in the antiferromagnetic phase, both magnons and domain-wall excitations that connect two parts of the chain with different antiferromagnetic order are relevant. 
Following Ref.~\cite{PerturbationTheoryD}, we call these domain-wall states holons and spinons. The spin configuration of a holon (spinon) state can be schematically written as $|...+-+\sigma -+-...\rangle$ with $\sigma = 0$ ($\sigma = \pm 1$), where  0 denotes a site with $S^z=0$ and $\pm$ a site with $S^z=\pm 1$. 
Holons are the lowest-lying excitations and their energy gap closes at the transition to the Haldane phase~\cite{PerturbationTheoryD}. 
Scattering between domain-wall excitations becomes important at finite temperature, similar to the Villain mode~\cite{Villain} in the antiferromagnetic phase of spin-1/2 chains. 
The MPS simulations take states with an odd number of domain walls into
account. In principle, it should be possible to exclude these states
from the calculation by adding a small staggered magnetic field. 
\begin{figure}[!bt]
\centering
\includegraphics[width=0.48\textwidth]{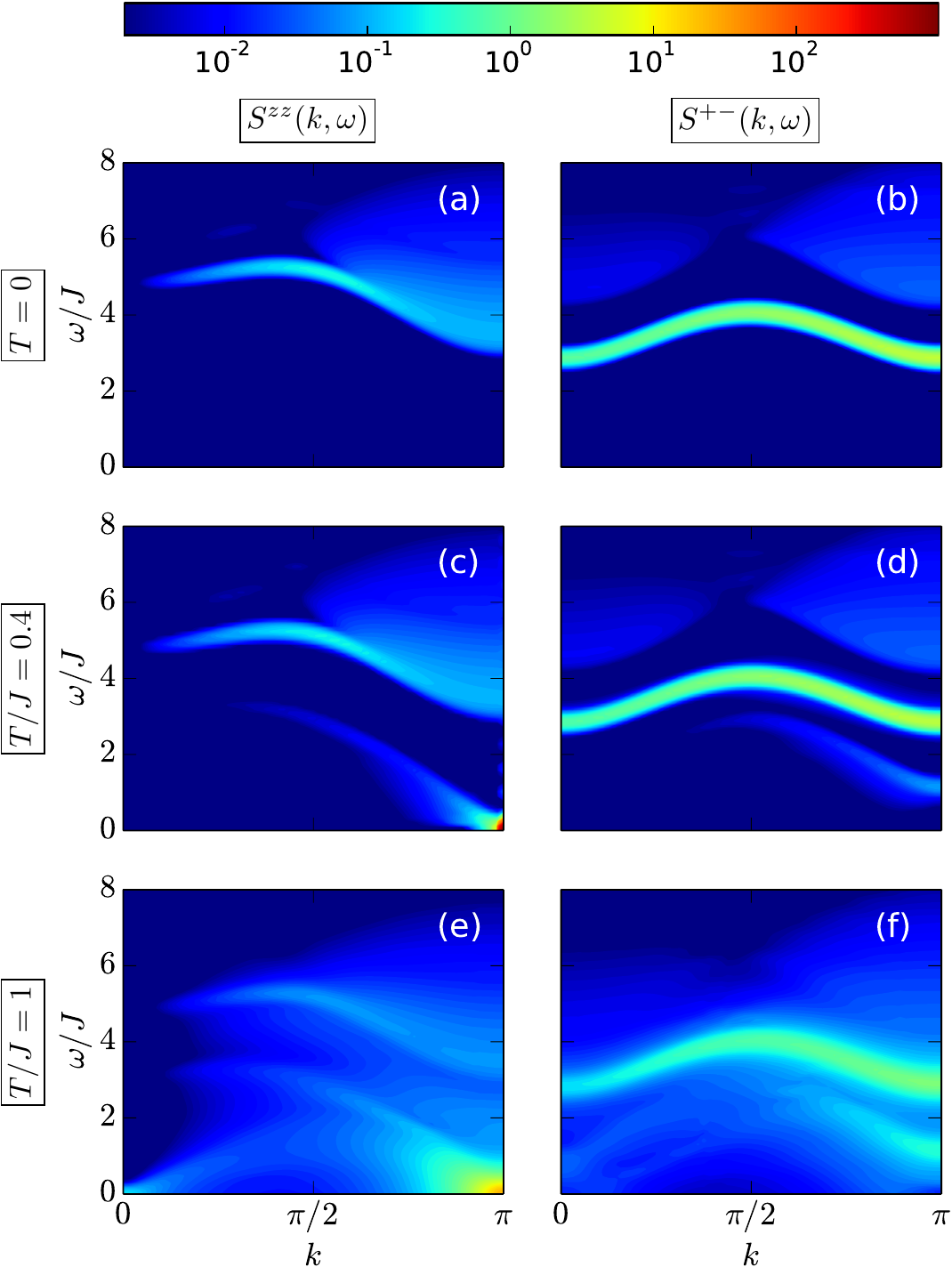}
\caption{Dynamic spin structure factor in the antiferromagnetic phase. Model parameters are $D/J=0.2$ and $J_z/J=2$. We use  $T/J=0,0.4$ and $1$ as in Fig.~\ref{SkwLargeD}.
 }
\label{SkwAFM}
\end{figure}
Figure~\ref{SkwAFM} shows the dynamic spin structure factors for $D/J=0.2$ and $J_z/J=2$.
In the zero-temperature longitudinal structure factor $S^{zz}(k,\omega)$, a bound state can be seen for $k\lesssim \pi/2$. It merges with the two-holon continuum for higher momenta. The small spectral weight above the bound state for $k\lesssim \pi/2$ corresponds to the two-spinon continuum. 
At finite temperatures, additional spectral weight shows up at low energies, which again can be related to intraband scattering. 
Holons are expected to provide the largest contribution. \fl{Most strikingly, the thermal intraband scattering leads to a strong increase of the longitudinal structure factor $S^{zz}(k,\omega)$ around $k=\pi$ and $\omega=0$ [see Fig.~\ref{SkwAFM}(c), and 4(e)]. }
When the temperature is increased, two separate peaks below the zero-temperature response become visible. From the dispersion relations of these excitations, one can deduce that the upper peak corresponds to holon intraband scattering and the lower one to either spinon or magnon intraband scattering. 
The transversal structure factor $S^{+-}(k,\omega)$ at zero temperature consists primarily of the single-magnon line and the spinon-holon continuum. 
Additional low-energy contributions occur for finite temperature. 
At low temperatures, the scattering between holons and spinons should be most significant. The momentum dependence of the spectral weight is weaker than for the intraband holon-scattering signal in $S^{zz}(k,\omega)$. 

To summarize, applying infinite boundary conditions to the time-dependent density-matrix renormalization group technique at finite temperatures, the dynamic spin structure factor has been analyzed for the Haldane, large-$D$ and antiferromagnetic (N\'eel) phases of the spin-1 $XXZ$ chain with onsite anisotropy. In each case, the finite-temperature result differs markedly from the one at zero temperature because of thermally activated scattering processes.  Our results reveal that further high-resolution inelastic neutron scattering experiments would be highly desirable to detect the thermally enhanced spectral weight and prove the differences in the magnetic response between the various spin-1 chain compounds.

\section*{Acknowledgements}
MPS simulations were performed using the ITensor 
library~\cite{ITensor}. 
F.L. was supported by Deutsche Forschungsgemeinschaft through Project No. FE 398/8-1.

\newpage
\appendix 
\section{\Large 
  Supplemental material}
\renewcommand{\theequation}{$\text{S}$\arabic{equation}}
\setcounter{equation}{0}
\renewcommand{\thefigure}{$\text{S}$\arabic{figure}}
\setcounter{figure}{0}

\setcounter{page}{1}
\makeatletter
\renewcommand{\bibnumfmt}[1]{[S#1]}
\renewcommand{\citenumfont}[1]{S#1}

\subsection{Constant-momentum cuts}
\begin{figure}[b!]
\centering
\includegraphics[scale=0.95]{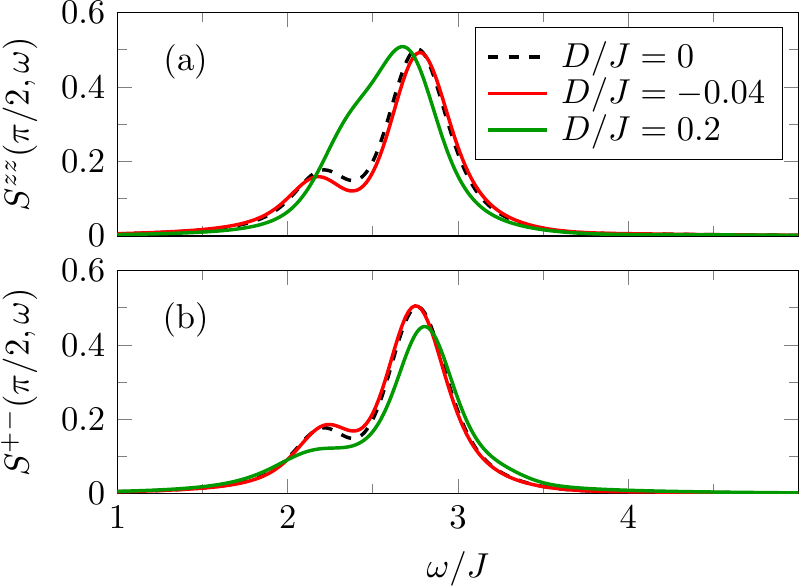}
\caption{Finite-temperature dynamic spin structure factor in the Haldane phase at fixed momentum $k=\pi/2$ and temperature $T/J=0.4$ for different values of the single-ion anisotropy $D$. Like in the main text, all spectral functions are convoluted with a Gaussian of width $0.1J$ and shown in units of $J^{-1}$.}
\label{figS1}
\end{figure}

To allow a better comparison with other theoretical results and experimental data, we plot the dynamic spin structure factor at selected constant momentum transfer. 
Fig.~\ref{figS1} shows the spectral functions in the Haldane phase for $k=\pi/2$ and temperature $T/J=0.4$, where both a coherent magnon mode and an intraband signal can be observed. 
For single-ion anisotropy $D/J=0.2$, the dynamic structure factor differs noticeably from the isotropic case. In particular, only a single peak appears in the longitudinal part $S^{zz}(k,\omega)$ [Fig.~\ref{figS1}(a)]. The effect of an anisotropy $D/J=-0.04$ is instead negligible. 

\begin{figure}[tb]
\centering
\includegraphics[scale=0.95]{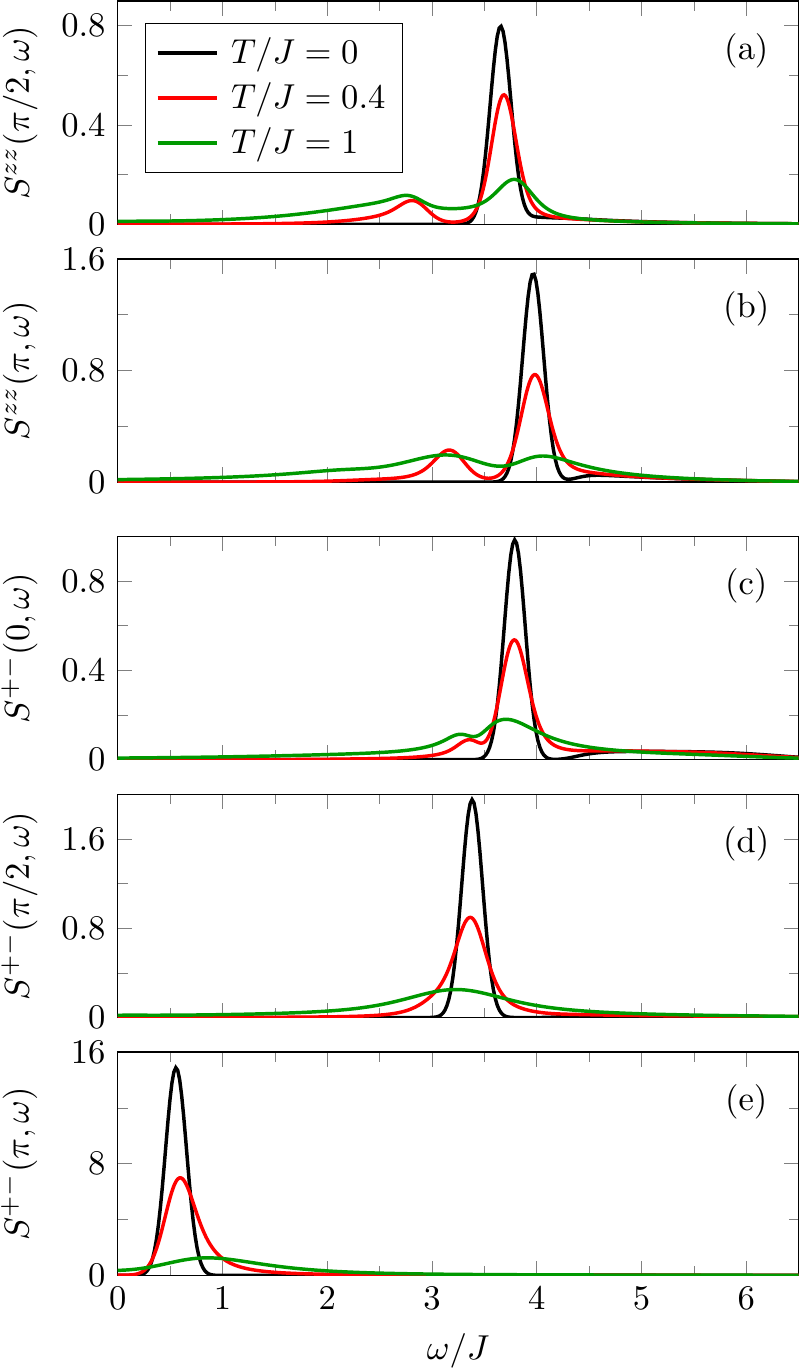}
\caption{Momentum cuts of the dynamic spin structure factor in the large-$D$ phase for $D/J=2$ and $J_z/J=1$.}
\label{figS2}
\end{figure}

Constant-momentum cuts of the dynamic structure factor in the large-$D$ phase are presented in Fig.~\ref{figS2}. At finite temperature, an intraband signal is clearly visible in the longitudinal response $S^{zz}(k,\omega)$ [panels (a) and (b)]. 
On the other hand, only a small additional peak  directly  below the exciton line shows up for small momentum in the transversal response $S^{+-}(k,\omega)$ [Fig.~\ref{figS2}(c)]. 
The thermal shift of the exciton mode in $S^{+-}(k,\omega)$ depends on the momentum. For $k=0$ and $k=\pi/2$, the peak moves to lower energy, while it moves to higher energy for $k=\pi$. 

\begin{figure}[t!]
\centering
\includegraphics[scale=0.95]{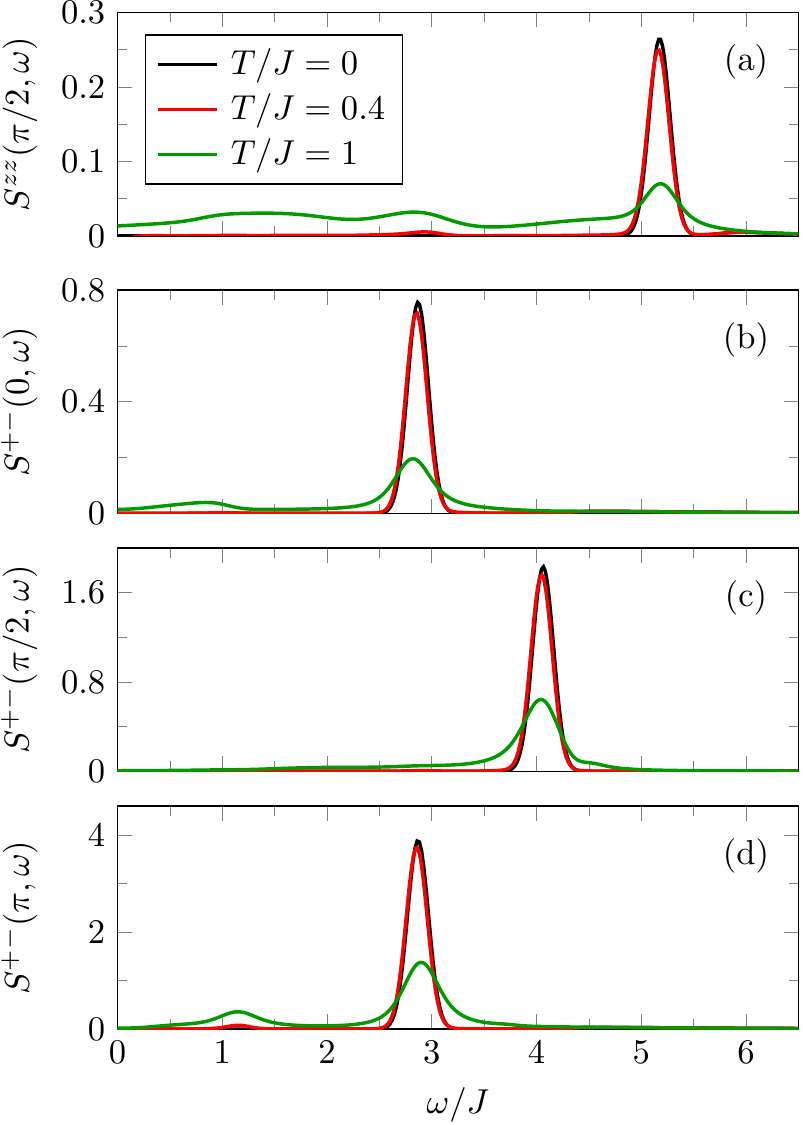}
\caption{Momentum cuts of the dynamic structure factor in the antiferromagnetic phase for $D/J=0.2$ and $J_z/J=2$.}
\label{figS3}
\end{figure}

Fig.~\ref{figS3} shows the dynamic structure factor for $J_z/J=2$ and $D/J=0.2$ in the antiferromagnetic phase. 
Because of the relatively large gap, the quasi-particle peaks are only weakly affected when the temperature is changed from $T=0$ to $T/J=0.4$. Further increasing the temperature to $T/J=1$, however, leads to a noticeable broadening. 
As in the other phases, low energy contributions to the structure factor show up at finite temperature that can be attributed to thermally activated scattering processes. In particular, we see two additional peaks in the longitudinal structure factor $S^{zz}(k,\omega)$ at momentum $k=\pi/2$ [Fig.~\ref{figS3}(a)]. 

\begin{figure}[t!]
\centering
\includegraphics[scale=0.88]{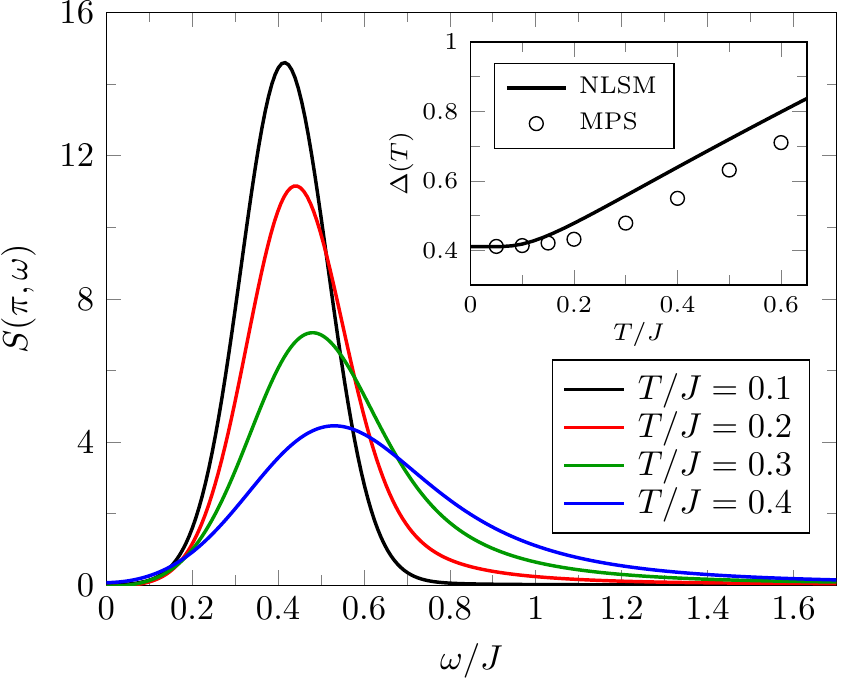}
\caption{Dynamic structure factor in the isotropic Heisenberg chain for momentum $k=\pi$ and several temperatures. The inset compares the position of the numerically obtained single-magnon line (open symbols) with a nonlinear sigma model prediction~\cite{PhysRevB.50.9265_supp}.}
\label{figS4}
\end{figure}

\subsection{Comparison with nonlinear sigma model predictions}
We now consider the isotropic Heisenberg chain ($J_z/J=1$, $D=0$) for which analytical predictions based on the nonlinear sigma model (NLSM) are available. 
Figure~\ref{figS4} shows the single-magnon peak in the dynamic spin structure factor at momentum $k=\pi$. 
When the temperature is increased, the magnon line shifts and broadens, and the line shape becomes asymmetric, with a larger spectral weight at high energies. Such an asymmetry has also been obtained in the O(3) NLSM~\cite{PhysRevB.78.100403_supp}. 
We compare the thermal shift of the magnon line in our numerical results with the NLSM calculation of Ref.~\cite{PhysRevB.50.9265_supp}. 
To take the asymmetric line shape into account, we define the line position as the energy corresponding to the full width at half maximum. 
Our results for the thermal shift agree qualitatively with the activated behaviour predicted by the NLSM. 
However, the NLSM description seems to overestimate the energy shift already for low temperatures. 
In contrast to the present results, a quantum Monte Carlo study~\cite{Goettinger_supp} has found an almost perfect agreement with the NLSM results for temperatures up to $T/J\approx 0.4$.

\end{document}